\documentclass[%
lengthcheck,
nofootinbib,
 amsmath,amssymb,
 aps,
 prl,
floatfix,
]{revtex4-2}

\usepackage{graphicx} % Include figure files
\usepackage{bm} % bold math
\usepackage[colorlinks,allcolors=blue]{hyperref}
\usepackage{microtype}
\usepackage{physics}

\def\equationautorefname#1#2\null{Eq.#1(#2\null)}

\renewcommand{\Im}{\operatorname{Im}}

\newcommand{\Jlor}{J_{\mathrm{lor}}}
\newcommand{\Jsm}{J_{\mathrm{sm}}}
\newcommand{\Hmod}{H_{\mathrm{mod}}}
\newcommand{\Jmod}{J_{\mathrm{mod}}}

\begin{document}

\title{A Lindblad master equation capable of describing hybrid quantum systems in the ultra-strong coupling regime}

\author{Maksim Lednev}
\author{Francisco J. García-Vidal}
\author{Johannes Feist}
\affiliation{Departamento de Física Teórica de la Materia Condensada and Condensed Matter Physics Center (IFIMAC), Universidad Autónoma de Madrid, E-28049 Madrid, Spain}

\begin{abstract}
Despite significant theoretical efforts devoted to studying the interaction between quantized light modes and matter, the so-called ultra-strong coupling regime still presents significant challenges for theoretical treatments and prevents the use of many common approximations. Here we demonstrate an approach that can describe the dynamics of hybrid quantum systems in any regime of interaction for an arbitrary electromagnetic (EM) environment. We extend a previous method developed for few-mode quantization of arbitrary systems to the case of ultrastrong light-matter coupling, and show that even such systems can be treated using a Lindblad master equation where decay operators act only on the photonic modes by ensuring that the effective spectral density of the EM environment is sufficiently suppressed at negative frequencies. We demonstrate the validity of our framework and show that it outperforms current state-of-the-art master equations for a simple model system, and then study a realistic nanoplasmonic setup where existing approaches cannot be applied.
\end{abstract}

\maketitle

Light-matter interaction in the strong coupling regime in which matter and electromagnetic (EM) modes hybridize has enabled the manipulation of the physical and chemical properties of hybrid light-matter systems at a quantum level~\cite{Garcia-Vidal2021,Kasprzak2006,Zhong2017,Chervy2016,Zasedatelev2021,Feist2015,Galego2016}. Typically, these are dissipative systems in which the interaction with an external reservoir introduces irreversible dynamics such as decay of excitations by photon emission. Such effects are customarily treated through Lindblad master equations, in which baths are represented by Lindblad dissipation terms~\cite{Carmichael1993Book,Breuer2007,Manzano2020}. These terms typically act on the uncoupled components to, e.g., represent losses of a cavity mode due to leakage through the mirrors. While it has long been known that decay operators derived for an uncoupled system can lead to unphysical effects in the coupled system~\cite{Carmichael1973}, their use often remains a reasonable approximation. However, this fails in the ultra-strong coupling (USC) regime where the coupling strength becomes a significant fraction of the system transition frequencies, which has been achieved in many physical systems ranging from organic molecules in Fabry-Pérot cavities to superconducting qubit-oscillator circuits~\cite{Askenazi2017,Yoshihara2017Superconducting,Yoshihara2017Characteristic,Yoshihara2018,Bayer2017,Maissen2014,Benz2016,Barachati2018Tunable,Gambino2014,Paravicini-Bagliani2019,Keller2020}. In this regime, the commonly used rotating-wave approximation for light-matter interaction breaks down, leading to entangled ground states with virtual excitations and opening new opportunities for nonlinear optics~\cite{Kockum2019, Forn-Diaz2019}. Decay operators in the uncoupled basis then introduce unphysical effects such as artificial emission from the ground state since they act on the virtual excitations~\cite{Ciuti2006,Beaudoin2011,Salmon2022}.

In order to mitigate these problems, decay operators acting in the coupled or dressed basis have been derived~\cite{Beaudoin2011,Ma2015,Bamba2014,Settineri2018}. However, the price to be paid is that the system Hamiltonian has to be diagonalized and that the decay operators become significantly more complex. The use of such approaches is thus restricted to simple cases where the dynamics is dominated by a single lossy cavity mode and the emitter has limited structure, while at the same time the use of few-mode and few-state approximations in the ultrastrong coupling regime becomes questionable~\cite{DeBernardis2018Breakdown,DiStefano2019,Saez-Blazquez2022Can}. Finally, these methods still employ an underlying Born-Markov approximation for the coupling of the system (emitter and discrete mode) to the outside environment and are in this respect similar to the standard Bloch-Redfield (BR) master equation of open quantum systems~\cite{Carmichael1993Book,Breuer2007}.

\begin{figure}[tb]
   \includegraphics[width=\linewidth]{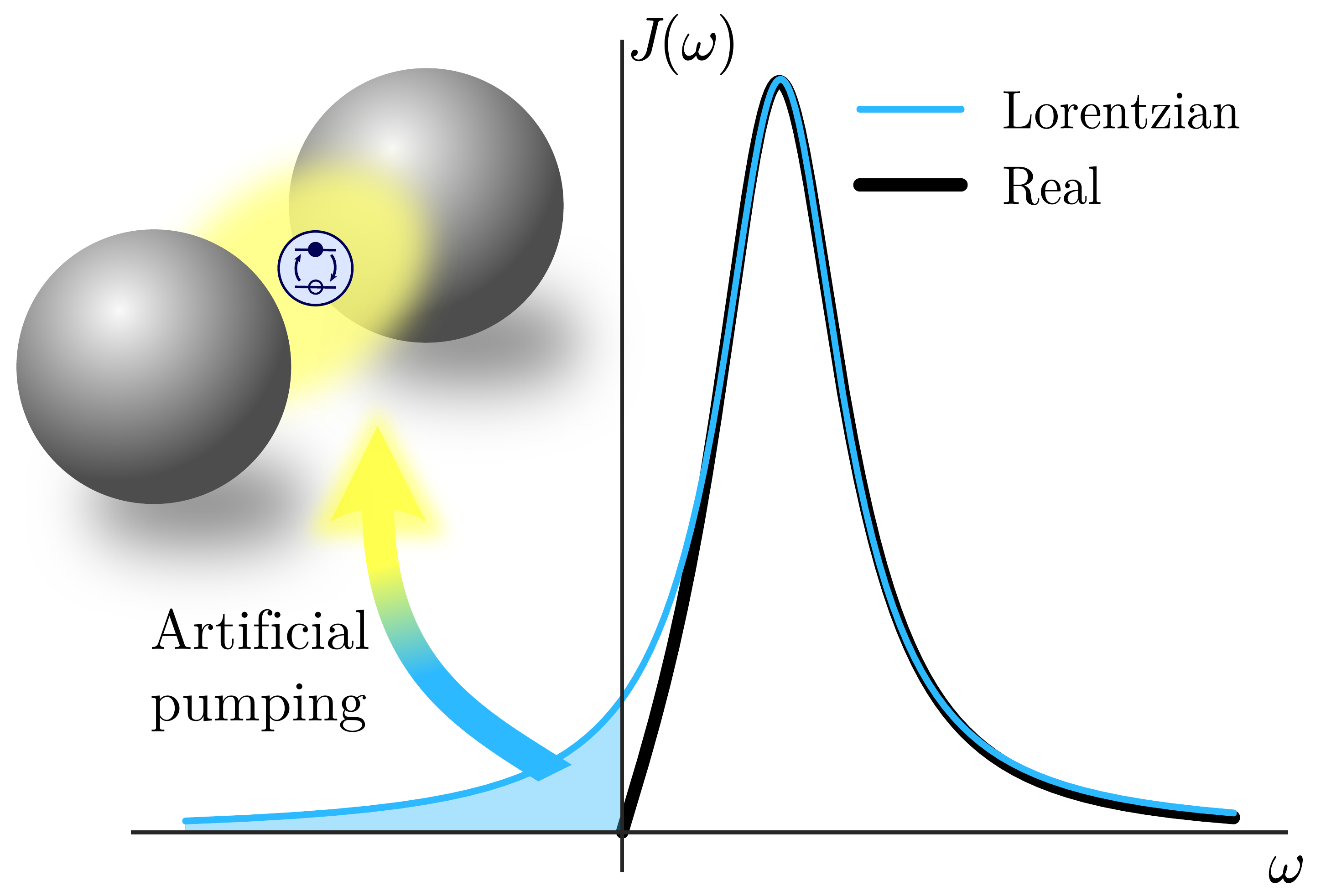}
   \caption{A realistic spectral density at zero temperature and its Lorentzian approximation, corresponding to a single-mode Lindblad master equation. The tail at negative frequencies causes artificial pumping of energy into the emitter, leading to incorrect dynamics.}
   \label{fig:sketch}
\end{figure}

In this Letter, we introduce a method for treating hybrid quantum systems in the USC regime that works for arbitrary EM environments, is independent of the properties of the quantum emitter, and does not make any Markovian approximation, while still maintaining the simple form of a Lindblad master equation in which the dissipators are simple decay terms for the ``cavity'' modes. We achieve this by extending a recently developed approach~\cite{Medina2021,Sanchez-Barquilla2022} to the case of extreme coupling between constituents or with external baths. The crucial step relies on the fact that the failure of the conventional approach can be understood from the perspective of open quantum systems theory~\cite{Breuer2007,Weiss2012,DeVega2017} by realizing that a single lossy cavity mode corresponds to an effective environment that contains negative-frequency components (see \autoref{fig:sketch}). We demonstrate that for an EM environment consisting of interacting modes~\cite{Medina2021,Sanchez-Barquilla2022}, their interference in the coupling to the emitter can suppress these negative-frequency components, making it possible to avoid the artificial effects inherent in the standard Lindblad master equation. We consider several limiting cases, benchmarking our approach by comparing with numerically exact solutions, and compare it to current state-of-the-art methods. Finally, we illustrate the capability of our method to go beyond simple ad-hoc model systems by considering the dynamics of an emitter placed in the hot spot of a nanoplasmonic structure formed by a dimer of silver spheres.

We start from the Hamiltonian describing the interaction of an emitter with its EM environment (with $\hbar=1$ here and below)~\cite{Buhmann2008Casimir,Feist2020}
\begin{equation}
    H = \frac{\omega_e}{2} \sigma_z + \int\limits_{-\infty}^{\infty} (\omega a_{\omega}^{\dagger} a_{\omega} + \sqrt{J(\omega)}(a_{\omega}^{\dagger} + a_{\omega})\sigma_x) \mathrm{d}\omega,
    \label{eq:full_H}
\end{equation}
where for simplicity we use a two-level emitter described by Pauli operators $\sigma_x$ and $\sigma_z$, with transition frequency $\omega_e$. The operators $a_{\omega}$ ($a_{\omega}^{\dagger}$) are bosonic annihilation (creation) operators for photon modes at frequency $\omega$. The spectral density $J(\omega)$ encodes the full information about the EM environment and its interaction with the emitter~\cite{Weiss2012,DeVega2017,Feist2020}.

We apply a recently developed few-mode quantization method based on replacing the EM system with an equivalent model consisting of $N$ lossy and interacting modes~\cite{Medina2021,Sanchez-Barquilla2022}. Its dynamics are given by
\begin{subequations}\label{eq:Hmodel}
   \begin{align}
      \Hmod &= \frac{\omega_e}{2} \sigma_z + \sum_{i,j} \omega_{ij} a_i^{\dagger} a_j + \sum_{i} g_i(a_i^{\dagger} + a_i)\sigma_x\\
      \dot{\rho} &= -i[\Hmod,\rho] + \sum_{i} \kappa_i L_{a_i}[\rho],
   \end{align}
\end{subequations}
where $L_{O}[\rho] = O\rho O^{\dagger} - \frac{1}{2}\{O^{\dagger}O,\rho\}$ is a Lindblad dissipator, $\omega_{ij}$ encodes the mode energies and couplings, $\kappa_i$ their decay rates, and $g_i$ their coupling to the emitter.
This model is exactly equivalent to \autoref{eq:full_H} with spectral density $\Jmod(\omega) = \frac{1}{\pi} \vec{g}\cdot \Im[(\tilde{\textbf{H}} - \omega)^{-1}] \cdot \vec{g}$, where $\tilde{\textbf{H}}_{ij} = \omega_{ij} - \frac{i}{2}\delta_{ij}\kappa_i$~\cite{Medina2021,Sanchez-Barquilla2022}. By varying parameters $\omega_{ij}$, $\kappa_i$, and $g_i$, the model can be adjusted to obtain a spectral density as close as desired to the original problem by performing a nonlinear fit of $\Jmod(\omega)$ to the physical spectral density $J(\omega)$ for a sufficient number of modes $N$. 

The single-mode case $N=1$ of \autoref{eq:Hmodel} is the conventional quantum Rabi model~\cite{Rabi1937,Braak2011}, and thus corresponds to a Lorentzian spectral density $\Jlor(\omega) = \frac{g^2}{\pi} \frac{\kappa/2}{(\omega_c-\omega)^2 + \kappa^2/4}$~\cite{Imamoglu1994,Tamascelli2018,Medina2021}, where $\omega_c$ is the mode frequency and $\kappa$ its decay rate. In contrast to physical spectral densities, $\Jlor(\omega)$ is non-zero along the whole real axis, including for negative frequencies. Since counterrotating coupling terms $a \sigma^-$, $a^\dagger \sigma^+$ are resonant at negative frequencies (where $\sigma^\pm$ are Pauli jump operators, with $\sigma_x = \sigma^+ + \sigma^-$), and the emission of negative-energy quanta corresponds to the absorption of energy by the system, this viewpoint provides a simple intuitive explication for the artificial pumping observed when cavity decay is described with a Lindblad term and the rotating-wave approximation is not performed. At the same time, it provides a natural recipe for preventing such effects: Ensuring that the spectral density at negative frequencies is sufficiently small. While this is not possible for noninteracting modes (corresponding to a sum of Lorentzians), we show below that interactions can enable destructive interference between the modes that allows this to be achieved even with a relatively small number of modes.

\begin{figure}[tb]
  \includegraphics[width=\linewidth]{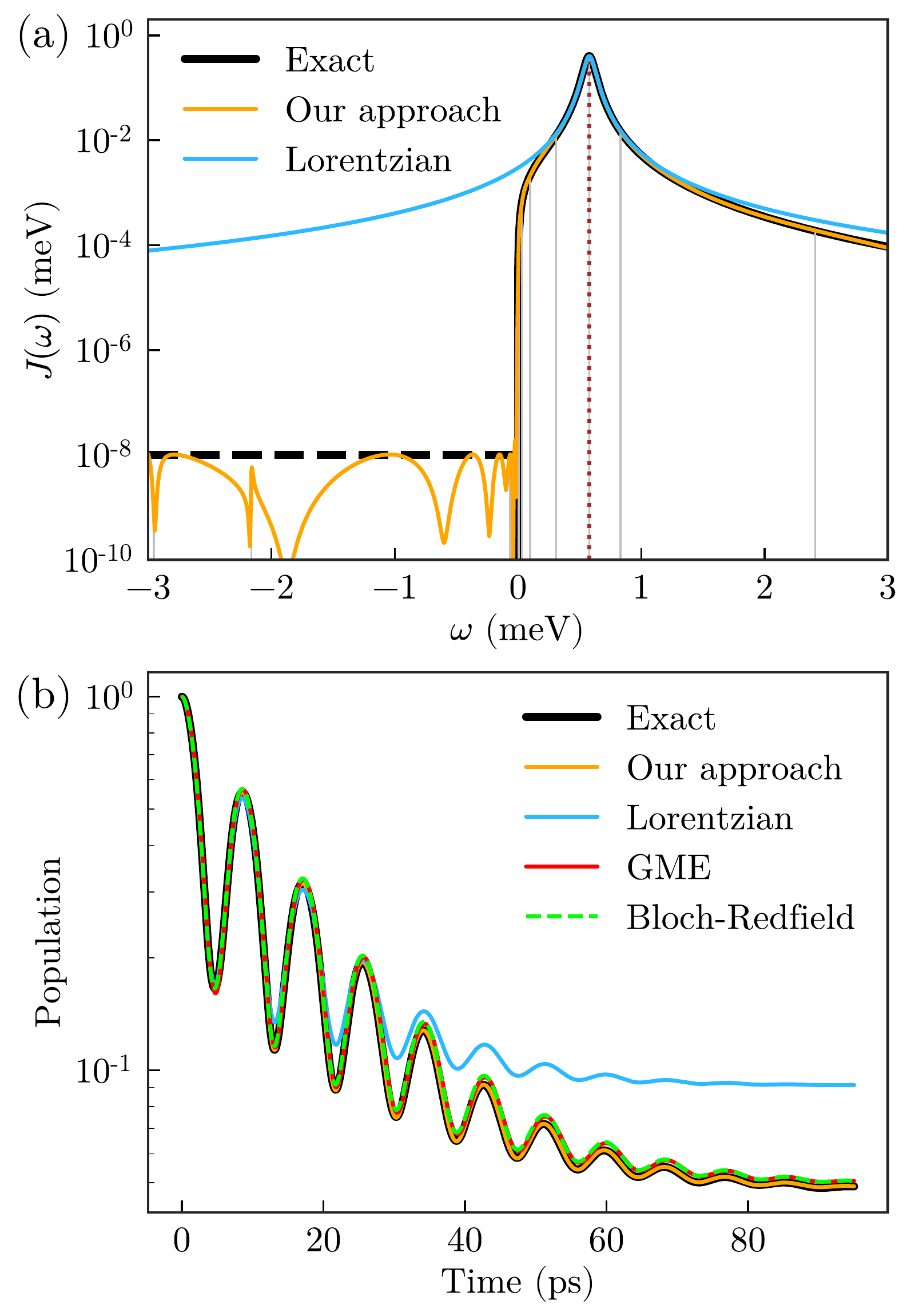}
  \caption{(a) The spectral density of the system under study. Black dashed line represents the preset threshold value for model spectral density at negative frequencies. Grey lines show real part of the complex resonances of the model spectral density, and the brown dotted line indicates the emitter transition frequency. (b) The population of the two-level emitter interacting with photonic environment in the USC regime for different methods.}
  \label{fig2}
\end{figure}

As a first test case, we study a single-mode setup corresponding to a physically allowed extension of the quantum Rabi model. This is obtained by coupling a single mode to an Ohmic ``background'' bath. The effective spectral density of the full EM environment can then be analytically obtained as
\begin{equation}
   \Jsm(\omega) = \theta(\omega) \frac{2g^2}{\pi}\frac{\kappa\omega_c\omega}{(\omega_c^2 - \omega^2)^2 + \kappa^2\omega^2},
   \label{J_analytic}
\end{equation}
where $\theta(\omega)$ is the Heaviside theta function. This expression fulfills the physical constraints for EM spectral densities: It only contains positive-frequency components and it tends to zero for $\omega\to 0$. We note that it can also be obtained as an antisymmetrized extension of the Lorentzian spectral density, i.e., $\Jsm(\omega) = \theta(\omega)(\Jlor(\omega) - \Jlor(-\omega))$ (with a small renormalization of the parameters). We choose parameters typical for Landau polaritons formed in semiconductor quantum wells in the USC regime~\cite{Forn-Diaz2019,Paravicini-Bagliani2019,Keller2020}, with values $\omega_c = \omega_e = 0.58$~meV, $g=0.25$~meV, and $\kappa=0.1$~meV. The corresponding spectral density is shown in \autoref{fig2}(a), together with fits using either a single Lorentzian or 10 interacting modes and including the additional suppression at negative frequencies. We compare various approaches to calculate the population dynamics of an initially excited emitter in \autoref{fig2}(b): (i) The single-mode Lindblad master equation, (ii) a BR master equation (for which the Ohmic bath coupling to the ``main'' mode is treated perturbatively), (iii) the generalized master equation (GME) introduced in Ref.~\cite{Settineri2018}, and (iv) our approach with a collection of interacting modes. As expected, the single-mode Lindblad equation significantly overestimates the steady-state population due to the presence of artificial pumping. Comparison with ``exact'' (i.e., numerically converged) results obtained through direct discretization demonstrates that our approach produces converged results within the linewidth of the figure. Both the GME and BR approaches produce similar results that are relatively close to the exact ones, but show significant deviations at later times. We attribute this to the Markov approximation for the coupling between the main cavity mode and its background bath inherent in these approaches.

\begin{figure}[tb]
  \includegraphics[width=\linewidth]{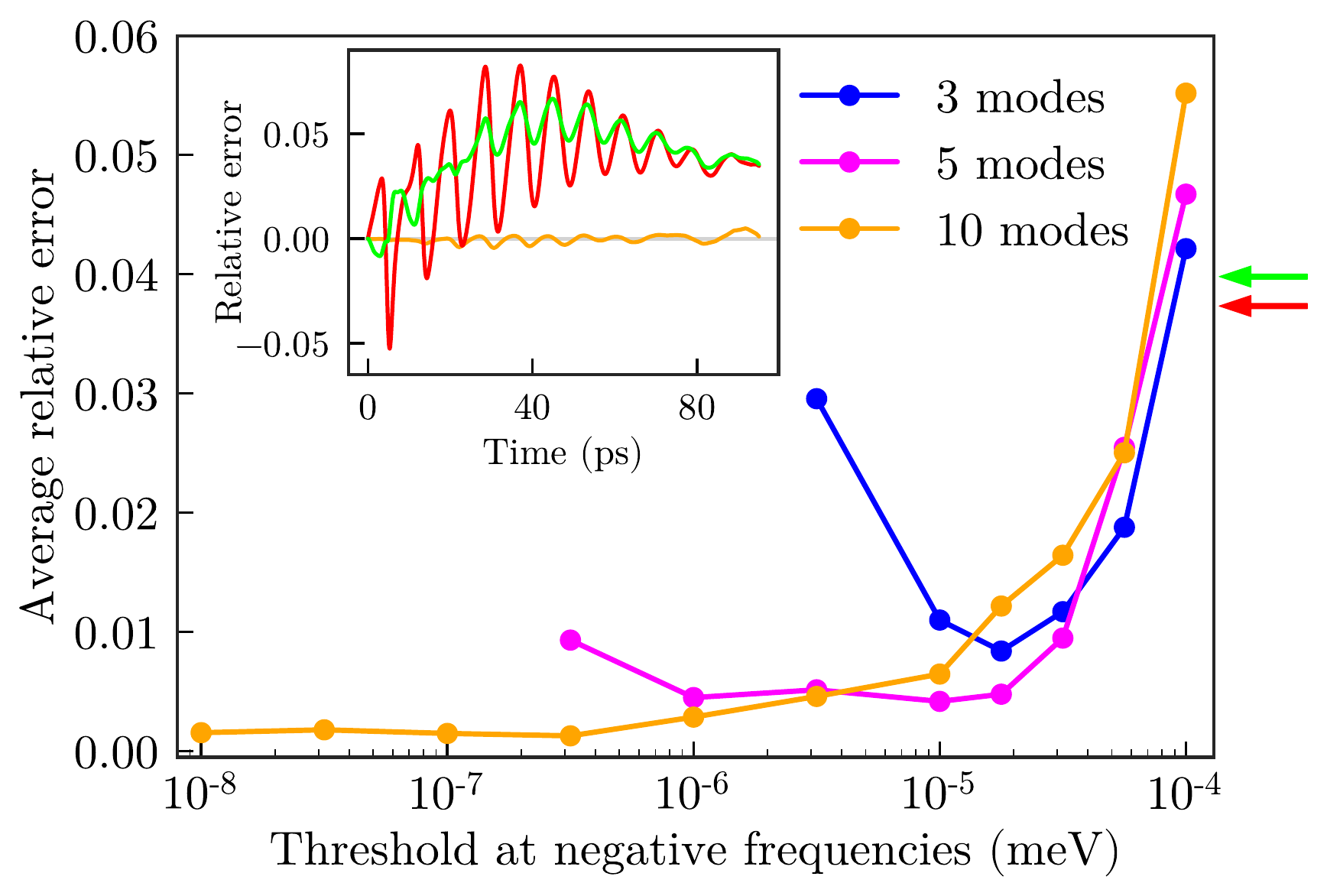}
  \caption{Average relative error of the emitter population dynamics obtained with our approach for different number of modes involved into fitting. Red and green arrows show the error of GME and BR master equations. In inset: time dependence of relative error. Orange, red and green lines represent result for our approach (with 10 modes at the threshold value = $10^{-8}$ meV), GME and BR master equation, respectively.}
  \label{fig3}
\end{figure}

In order to quantify the accuracy of the different methods, in \autoref{fig3} we show the relative error in the emitter population (compared to the ``exact'' numerical solution) for the different approaches. The BR and GME methods contain no free parameters and thus provide no way to systematically improve the approximation. Their time-dependent relative error in the emitter population is shown in the inset of \autoref{fig3}, with its value averaged over the propagation time indicated by the arrows in the main figure. In contrast, the fitting-based approach introduced here permits to choose a tradeoff between accuracy and complexity of the model by choosing the number of modes as well as the threshold value below which we require the spectral density at negative frequencies to be suppressed. The behavior of the method is demonstrated for the cases of $N=3$, $N=5$, and $N=10$ modes. When the threshold is set relatively high, the error is dominated by the artificial pumping and somewhat independent of the number of modes used for fitting, although we note that even for the largest threshold values considered here, the errors in our approach are comparable to the ones of the GME and BR methods. Decreasing the threshold initially leads to a significant reduction in the error for any number of modes. However, the error increases again at some point as the threshold is decreased, essentially because the fit quality in the positive-frequency components cannot be maintained when the constraints at negative frequencies are too stringent. This can be mitigated by increasing the number of modes in the fit, with $N=10$ modes providing enough flexibility in the current case to maintain accuracy even at the smallest threshold values we set ($J(\omega) \leq 10^{-8}$~meV for $\omega<0$). We note that even for 3 modes, the optimal accuracy of our approach is considerably higher than that of the BR and GME methods, while providing the additional advantage that the dissipators in the Lindblad master equation act only on the cavity modes and no diagonalization of the full Hamiltonian is required.

Next, we demonstrate the implementation of our model in a nanophotonic structure with a more complex spectral density: a dimer of silver spheres (radius 15~nm) with a 1~nm gap between them, embedded in a GaP matrix ($\epsilon=9$). We study the excited state population of a two-level quantum emitter placed in the center of the gap between the spheres. The emitter has a transition frequency $\omega_e = 2.4$~eV and dipole moment $\mu=20$~D oriented along the line connecting the centers of the spheres, which are typical parameters for organic molecules or quantum dots. Using the boundary element method implemented in SCUFF-EM~\cite{Reid2015}, we computed the dyadic EM Green function of the structure, which determines its spectral density~\cite{Feist2020}.

\begin{figure}[t]
   \includegraphics[width=\linewidth]{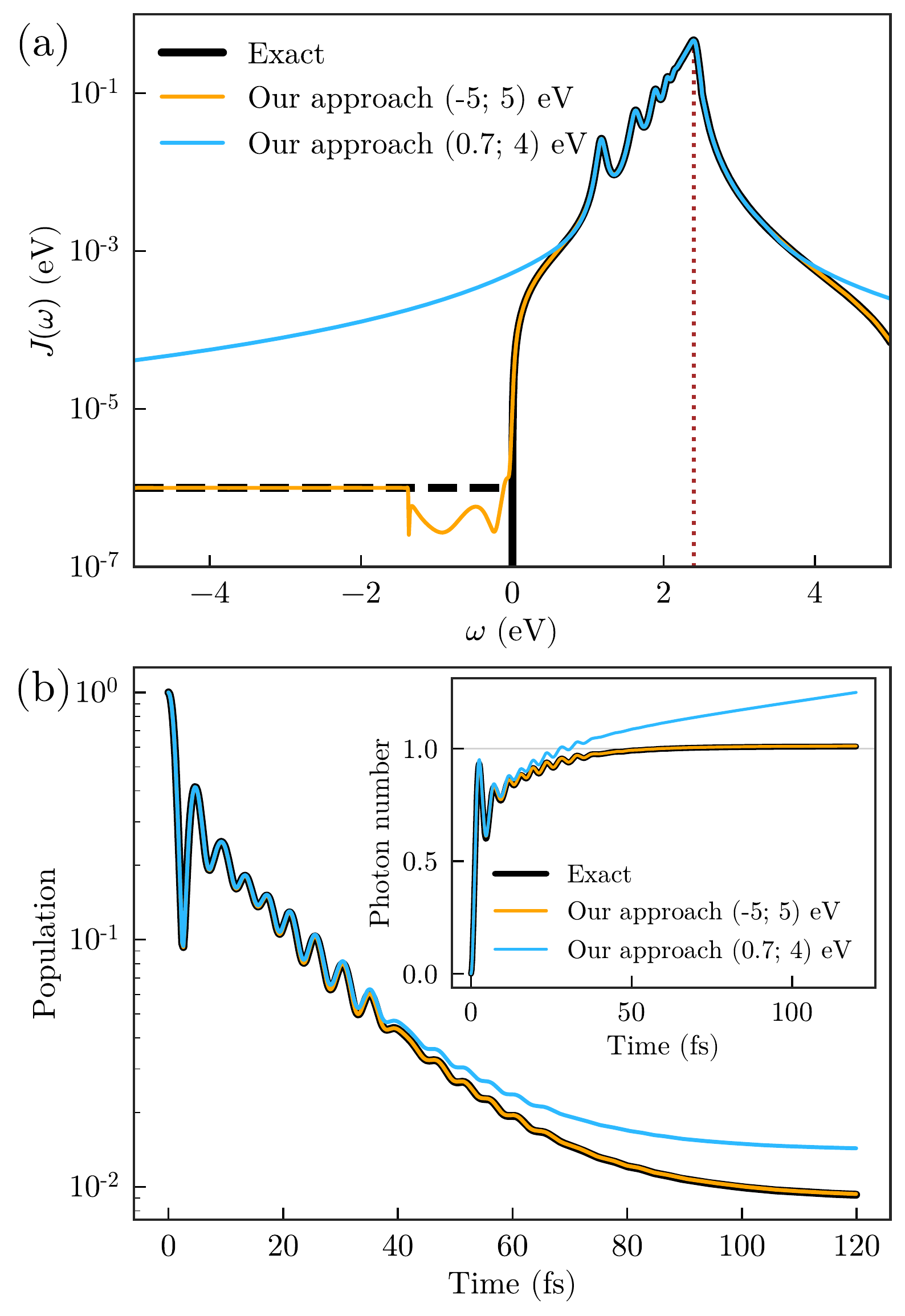}
   \caption{(a) The spectral density of the considered system. Black dashed line represents the preset threshold value for model spectral density at negative frequencies and the brown dotted line indicates the emitter transition frequency. (b) Emitter population as a function of time for numerically exact and our approach applied for narrow ($(0.7; 4)$~eV) and broad ($(-5; 5)$~eV) spectral windows. In inset: The total number of photons accumulated in the system and bath modes.}
   \label{fig4}
\end{figure}

The spectral density, shown in \autoref{fig4}(a), consists of several localized surface plasmon resonances of various multipolar orders. This already implies that the number of required modes for an accurate fit will be significantly larger than in the simplified models treated up to now. The high-permittivity dielectric background red-shifts these resonances compared to the free-space situation, making the fitting procedure for our approach more challenging since the target spectral density shows a significant gradient close to zero frequency. In order to obtain an accurate fit for the spectral density with sufficient suppression at negative frequencies, we used 28 modes. Since in this case, it is not possible to compare with simplified master equations or single-mode models, we instead compare with a fit where the negative frequency range is ignored and only the resonance peaks (in the frequency range between $0.7$~eV and $4$~eV) are fitted. For this simpler situation, the fit is already well-converged with 12 modes.

The emitter frequency $\omega_e$ is resonant with the pseudomode formed by overlapping high-order surface plasmon modes~\cite{Delga2014,Li2016Transformation}, i.e., close to the peak of the spectral density. We show the emitter population after initial excitation in \autoref{fig4}(b). After several oscillations due to interaction with the complex EM environment, it gradually decreases to zero while undergoing several further oscillations. Our method again provides essentially perfect agreement with the numerically exact discretization method. Moreover, we note that it is essential to include the suppression of the spectral density at negative frequencies to obtain converged results when counter-rotating coupling terms are included. The fit taking into account only the positive-frequency region from $0.7$ to $4$~eV leads to considerable deviations in the dynamics. As in the previous cases, the differences between the results of our approach for broad and narrow spectral ranges are caused by artificial pumping and, notably, are most pronounced at long times. In contrast, the oscillatory dynamics at short times is represented correctly in both cases since it is principally determined by the interaction between the emitter and the resonant region of the spectral density.

In addition to emitter population, in the inset of \autoref{fig4}(b), we show the time-dependent total number of photons in the system and bath modes. The exact dynamics demonstrates that the steady-state photon number slightly exceeds unity. This is a consequence of the importance of the counter-rotating terms, which lead to the (virtual) excitation of the cavity modes in the coupled system even in the ground state. Since the decay operators in our approach are of a Lindblad form, the population of the bath modes can be easily tracked as $P_{bath}(t) = \sum_{i=1}^N\int_0^{t}\kappa_i\langle a_i^{\dagger} a_i\rangle(t') dt'$. If the fit is performed only at positive frequencies without taking the suppression at negative frequencies into account, this results in a linear increase in the photon number at long times due to the continued artificial pumping through emission of (negative-frequency) photons from the ground state. In contrast, the result of our approach for the extended spectral region perfectly follows the trend of the numerically exact results. This shows that even though the discrete modes in the fit are in some sense arbitrary, their combination not only provides the correct emitter dynamics, but also correctly reproduces the temporal behavior of the photon number. 

We furthermore note that the steady state reached by the hybrid system becomes essentially pure when the artificial pumping is sufficiently suppressed. In the case of the extended spectral range, the steady-state density matrix $\rho_{s}$ is highly pure, $1 - \Tr(\rho_{s}^2) < 10^{-3}$, meaning that $\rho_{s}$ can be expressed to a good approximation as describing a pure quantum state $\rho_{s} \approx \ket{\psi_0}\bra{\psi_0}$. Furthermore, $\ket{\psi_0}$ corresponds to the eigenstate of the Hamiltonian with the smallest number of excitations. This is a consequence of the fact that the steady state is simply the ground state of the full system, which is unaffected by the decay terms in the master equation. The steady state could thus be obtained from just the Hamiltonian without having to invoke the Lindblad master equation.

To conclude, we have demonstrated a powerful method able to reproduce dynamics of a quantum system interacting with an arbitrary EM environment in any coupling regime by exploiting the mapping between nanophotonic spectral densities and Lindblad-form master equations~\cite{Medina2021,Sanchez-Barquilla2022}. By suppressing the spectral density at negative frequencies, artificial effects inherent to standard Lindblad forms can be removed. For simple model systems, we showed that the accuracy of our approach considerably exceeds the one of state-of-the art master equations. Furthermore, the approach can deal with realistic nanophotonic systems where simple models are not available. It thus offers a straightforward way for investigating extreme regimes of light-matter interaction while taking into account the full mode spectrum of complex EM environments.

\begin{acknowledgments}
We acknowledge support by the European Research Council through Grant ERC-2016-STG-714870, by the Spanish Ministry for Science, Innovation, and Universities -- Agencia Estatal de Investigación (AEI) through Grants PID2021-125894NB-I00 and CEX2018-000805-M (through the María de Maeztu program for Units of Excellence in Research and Development), by the Comunidad de Madrid through Proyecto Sinérgico CAM 2020 Y2020/TCS-6545 (NanoQuCo-CM) and by the ``(MAD2D-CM)-UAM7'' project funded by the Comunidad de Madrid, by the Recovery, Transformation and Resilience Plan from Spain, and by NextGenerationEU from the European Union.
\end{acknowledgments}

\bibliography{references}
\end{document}